\documentclass[%
 reprint,
 amsmath,amssymb,
 aps,
]{revtex4-2}

\usepackage{graphicx}
\usepackage{dcolumn}
\usepackage{bm}
\usepackage{xcolor}

\newcommand{\be}{\begin{equation}}
\newcommand{\ee}{\end{equation}}
\newcommand{\ba}{\begin{array}}
\newcommand{\ea}{\end{array}}
\newcommand{\bea}{\begin{eqnarray}}
\newcommand{\eea}{\end{eqnarray}}

\def\anu{{\bar\nu}}

\newcommand{\beq}{\begin{equation}}
\newcommand{\eeq}{\end{equation}}
\newcommand{\beqa}{\begin{eqnarray}}
\newcommand{\eeqa}{\end{eqnarray}}

\newcommand{\tx}{{\theta_{12}}}
\newcommand{\ty}{{\theta_{13}}}
\newcommand{\tz}{{\theta_{23}}}

\newcommand{\dl}{{\Delta_{31}}}
\newcommand{\ds}{{\Delta_{21}}}

\newcommand{\dcp}{\delta_{\mathrm{CP}}}
\newcommand{\nova}{NO$\nu$A~}

\newcommand{\pme}{P_{\mu e}}

\newcommand{\pmebar}{P_{\bar{\mu} \bar{e}}}

\newcommand{\dchsq}{\Delta\chi^2}

\begin{document}

\preprint{APS/123-QED}

\title{Update on non-unitary mixing in the recent \nova and T2K data}
\author{Xin Yue Yu}
\author{Zishen Guan}
\author{Ushak Rahaman}
\email{ushak.rahaman@cern.ch}
\author{Nikolina Ilic}
 
\affiliation{%
 Department of Physics, University of Toronto, Toronto, ON M5S 1A7, Canada
}%




\date{\today}

\begin{abstract}
In this paper, we have tested the non-unitary mixing hypothesis with the latest data from \nova and T2K experiments. We have also analysed their combined data. We have provided the best-fit values of the standard and non standard parameters after the analysis. $90\%$ limits on the non-unitary mixing parameters have also been provided. The constraints on unitary violation is stronger, compared to the constraints obtained from previous data from \nova and T2K. The tension between \nova and T2K at the $1\,\sigma$ for normal mass hierarchy can be reduced for non-unitary mixing due to $\alpha_{10}$, albeit for a value of $|\alpha_{10}|$ larger than the present global $90\%$ limit. Additionally a study of the future sensitivity of NO$\nu$A, T2K and DUNE has been provided.
\end{abstract}

\maketitle

\section{Introduction}
\label{ch:introduction}
 The neutrino oscillation phenomenon, driven by three mixing angles $\tx$, $\ty$ and $\tz$; two mass squared differences $\ds=m_{2}^2-m_{1}^{2}$ and $\dl=m_{3}^{2}-m_{1}^2$, where $m_i$s are the absolute masses of three neutrino mass eigen states $\nu_i$s, with $i=1,2,3$; and a CP violating phase $\dcp$, provides one of the windows to physics beyond the standard model (BSM). The currently unknown properties related to neutrino oscillation physics are the sign of $\dl$, octant of $\tz$, and the value of $\dcp$. Depending on the sign of $\dl$, there can be two different mass hierarchies: normal hierarchy (NH) for $\dl>0$; and inverted hierarchy (IH) for $\dl<0$. Similarly, if $\sin^22\tz<1$, there can be two different octants of $\tz$: lower octant (LO) for $\tz<\pi/4$; and a higher octant (HO) for $\tz>\pi/4$. The present long-baseline accelerator neutrino experiments \nova \cite{NOvA:2007rmc} and T2K \cite{T2K:2001wmr} are expected to measure these unknowns. However, the 2020 and 2024 data from \nova \cite{NOvA:2021nfi, Wolcott:2024} is in mild tension \cite{Rahaman:2021zzm} with the latest T2K data from 2020, \cite{T2Kapp, T2Kdisapp}  for the the $\dcp$ measurements and both experiments disfavour each other's $1\,\sigma$ allowed regions on the $\sin^2\tz-\dcp$ plane. These tensions opened up the possibility of the existence of BSM physics in the \nova and T2K data \cite{Rahaman:2021leu, Miranda:2019ynh, Chatterjee:2020kkm, Rahaman:2022rfp, Chatterjee:2024kbn, Denton:2020uda}. The recent joint analysis of \nova and T2K collaborations also observed this tension \cite{T2K:2025wet}. We have presented our analysis of the latest \nova and T2K data in Appendix~\ref{uni-analysis}. In this paper, we explore the non-unitary mixing in the \nova and T2K experiment. This is an update from ref.~\cite{Miranda:2019ynh}. Here, we consider one non-unitary parameter at a time, unlike the referenced analysis where all of the parameters simultaneously analyzed. This has allowed us to pinpoint the exact effects of non-unitary parameters on oscillation probabilities and event numbers. Best-fit values of standard oscillation parameters as well as non-unitary parameters have been presented. A $90\%$ limit on the non-unitary parameters have also been obtained from the present \nova and T2K data.  We also provide a theoretical explanation of our results, based on the effects of different parameters on the oscillation probabilities. Finally, we consider the role of a future combined result of \nova and T2K, and the upcoming long-baseline experiment, DUNE \cite{Abi:2018dnh}, under the assumption that non-unitary mixing exists.

 In section \ref{ch:non-uni}, we have introduced non-unitary mixing and discussed how it can arise in neutrino oscillation experiments. We have presented the results of our data analysis in section \ref{ch:results}. The results from future sensitivity studies have been presented in section~\ref{ch:future}, and the final conclusions have been drawn in section \ref{ch:conclusion}.

\section{Non-unitary mixing}
\label{ch:non-uni}
If more than three neutrino generations exist as iso-singlet heavy neutral leptons (HNL), they would not take part in neutrino oscillations in the minimal extension of the standard model. However, their ad-mixture in charged current weak interactions will affect neutrino oscillation and the neutrino oscillation will be described by an effective $3\times3$ non-unitary mixing matrix. In case of non-unitary mixing, the effective $3\times 3$ mixing matrix can be written as \cite{Xing:2007zj, Escrihuela:2015wra}:
\begin{equation}
N=N_{NP}U_{3\times 3}= \left[ {\begin{array}{ccc}
   \alpha_{00} & 0 & 0 \\
   \alpha_{10} & \alpha_{11} & 0 \\
   \alpha_{20} & \alpha_{21} & \alpha_{22}
  \end{array} } \right] U_{\rm PMNS} \, 
  \label{alpha-mat}
\end{equation}
where $U_{\rm PMNS}$ is the standard $3\times 3$ PMNS mixing matrix. The diagonal elements $\alpha_{ii}$ of $N_{NP}$ are real, and the off-diagonal elements $\alpha_{ij}=|\alpha_{ij}|e^{i\phi_{ij}}$ are complex, with $i,j=1,2,3$ and $i>j$. The details of the calculation of the oscillation probability with non-unitary mixing have been discussed in ref.~\cite{Miranda:2019ynh}. The present $3\,\sigma$ boundary values for non-unitary parameters are given in ref.~\cite{Escrihuela:2016ube, Blennow:2023mqx}. From ref.~\cite{Blennow:2023mqx}, it is clear that there is a stringent constraint on non-unitary neutrino mixing from the charged lepton flavour violation (CLFV) experiments. However, it is possible to obtain percent level non-unitary mixing in neutrino oscillation without violating the strong constraints of CLFV experiments in certain neutrino mass models involving low scale typw-I seesaw mechanisms, namely inverse and linear seesaw \cite{Forero:2011pc}. Besides it is important to independently test unitary violation of the three light neutrino mixing in the neutrino oscillation experiments. Several other works \cite{KM3NeT:2025ftj, Kozynets:2024xgt, Gariazzo:2022evs, Agarwalla:2021owd, Rahaman:2021cgc, Forero:2021azc, Ellis:2020hus} have studied unitary violation in the future simulated neutrino oscillation experiments as well as the present neutrino oscillation experiment data. If there is a mismatch between the CLFV results, and the neutrino oscillation results, that should be something to ponder about.

The low scale see-saw mechanism can allow percentage level deviation from unitary mixing at the CLFV experiments only for channels involving tau decay because of weaker constraints on lepton flavour violating tau decays. However, it is possible to evade all the constraints on non-unitery mixing from CLFV data and invoke large non-unitary mixing in neutrino oscillation experiments in another way. If the sterile neutrino is light enough to be produced in all relevant processes including $\beta$-decay, then unitarity is preserved in CLFV experiments because all the mass eigenstates are kinemetically accessible. However, the sterile neutrinos will mix and propagate with 3 standard neutrinos and hence participate in the oscillation phenomenon. Thus, their effects will be observed in neutrino oscillation experiments. If the additional mass eigen states are light enough, they can be directly probed in neutrino oscillation experiments \cite{Kopp:2013vaa}. However, if the new mass eigenstate lead to large $\Delta m^2L/E$, the oscillation will average out at the detector. In this averaged out region, the phenomenology of sterile neutrino is equivalent to that of non-unitary mixing \cite{Blennow:2016jkn, Fong:2016yyh}, except a sub-leading constant term. Moreover, the zero-distance effect \cite{Blennow:2016jkn} for non-unitary mixing will not be observed because the sterile oscillation would not develop yet at the near detector. In this particular case, it is possible to observe non-unitary mixing effect in neutrino oscillation experiments without observing them in CLFV experiments. Ref.~\cite{Blennow:2025qgd} have discussed this in details. In case of a $3+1$ scenario, it is possible to draw a direct correspondence between the non-unitary mixing parameters in eq.~\ref{alpha-mat} and the parameters of the sterile neutrino mixing in the complete $4\times 4$ mixing matrix. Moreover, from table 1 of ref.~\cite{Blennow:2025qgd}, it is obvious that for this scenario, it is possible to see large effect of non-unitary mixing in neutrino oscillation experiments without violating CLFV constraints.

In our analysis, we have considered $\alpha_{00}$, $\alpha_{10}$, and $\alpha_{11}$ as the possible source of the non-unitary effect, since these three parameters have the maximum effect on $\pme$ and $\pmebar$, which are the oscillation probabilities for $\nu_e$ and $\bar{\nu_e}$ appearances from a $\nu_\mu$ beam. The details of our analysis are provided in Appendix \ref{analysis}.

\section{Results} 
\label{ch:results}
\begin{figure}[h!]
\centering
\includegraphics[width=0.5\textwidth] {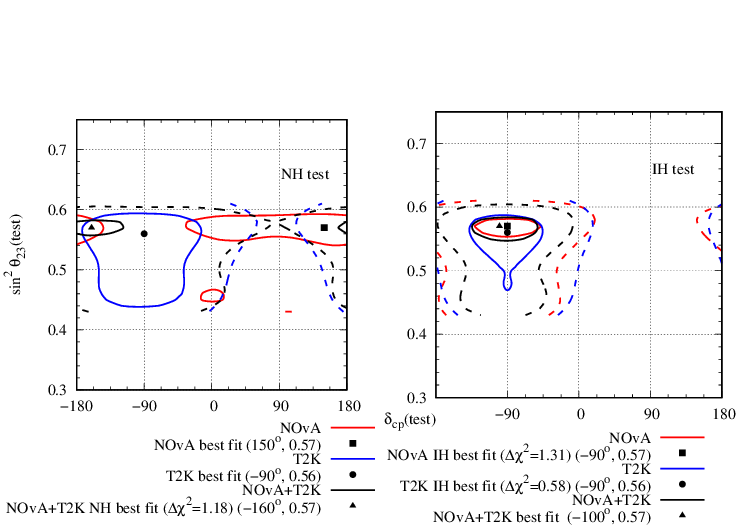}
\includegraphics[width=0.5\textwidth] {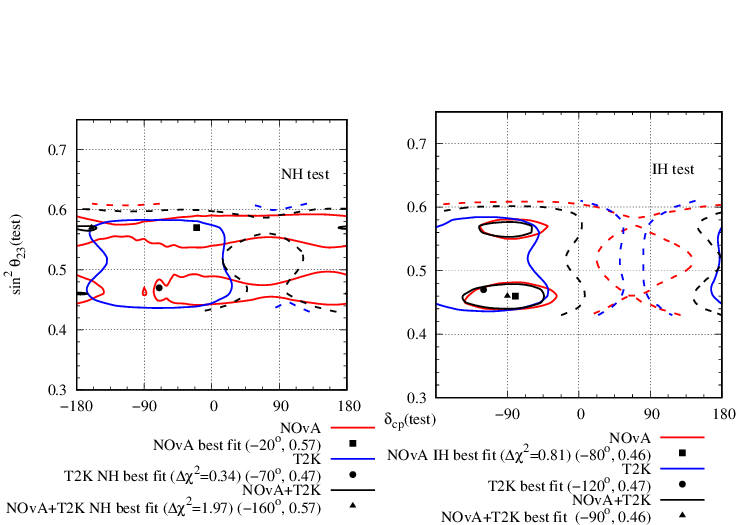}

\caption{\footnotesize{Allowed regions in the $\sin^2\tz-\dcp$ plane for \nova and T2K after analysing the data with non-unitary mixing with $\alpha_{00}$ ($\alpha_{10}$) in the upper (lower) panel. The left (right) panel is for NH (IH). The red (blue) line indicates \nova (T2K), and the black line indicates the combined data. The solid (dotted) lines indicate the boundaries of the $1\,\sigma$ ($3\,\sigma$) allowed regions.}}
\label{result1}
\end{figure}

From Fig.~\ref{result1} it can be seen that for non-unitary mixing arising due to $\alpha_{00}$, the two experiments have a little overlap region at a $1\,\sigma$ confidence level (C.L.) for NH. However, the overall characteristic is similar to that of the results obtained by analysing the data with standard unitary mixing (see fig.~\ref{result-uni}).

For $\alpha_{10}$, the $1\, \sigma$ overlap between two experiments for NH is larger. As in the preceding case, \nova loses its $\dcp$ sensitivity for NH. The T2K best-fit point occurs at the IH and with $\tz$ in the LO. However, there exist degenerate best-fit points at IH-HO ($\dchsq=0.74$), NH-HO ($\dchsq=0.72$), and NH-LO ($\dchsq=0.34$). 

We also did similar analysis with non-unitary mixing due to $\alpha_{11}$. The result on $\sin^2\tz-\dcp$ plane is quite similar to the results for $\alpha_{00}$. That is why we have not shown the results on $\sin^2\tz-\dcp$ plane with $\alpha_{11}$ as the source of unitary violation here.

\begin{figure}[h!]
\centering
\includegraphics[width=0.4\textwidth] {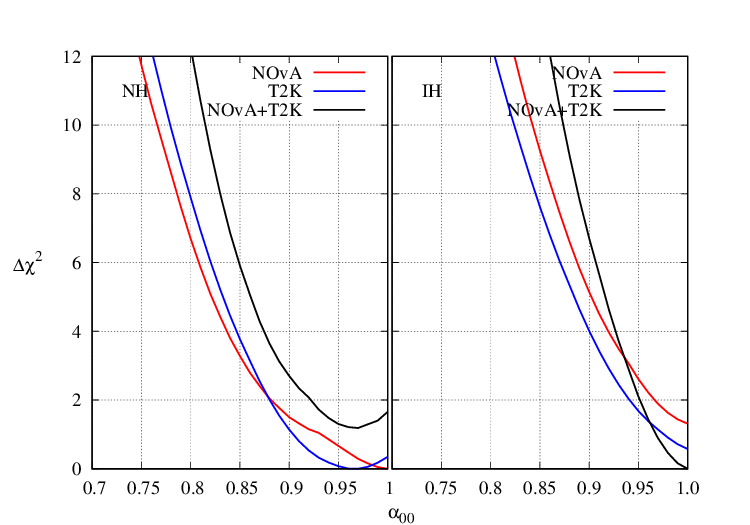}
\includegraphics[width=0.4\textwidth] {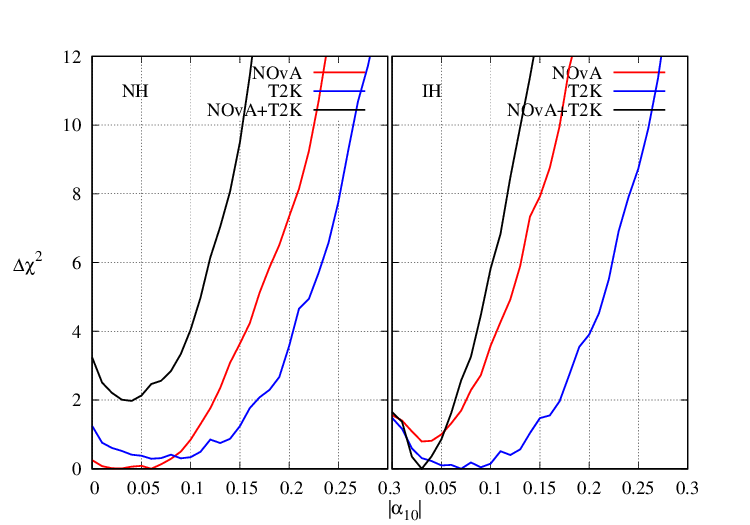}
\includegraphics[width=0.4\textwidth] {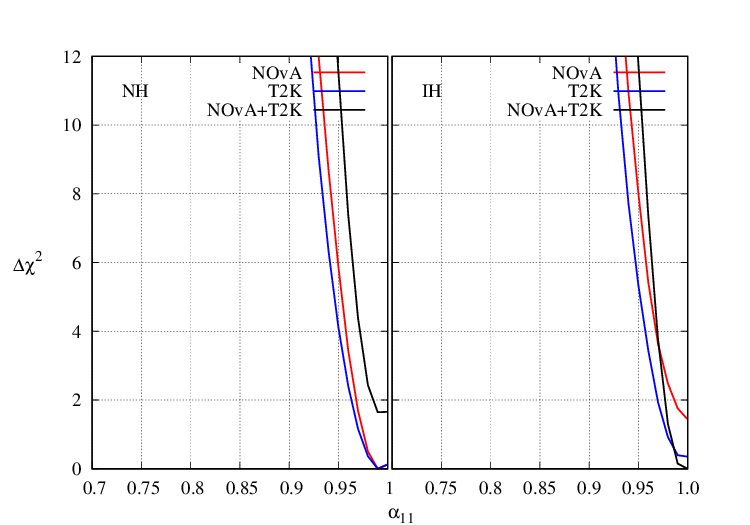}
\caption{\footnotesize{$\dchsq$ as a function of individual non-unitary parameters for 2024 long baseline data.}}
\label{result2}
\end{figure}

From Fig.~\ref{result2}, it can be observed that for the $\alpha_{00}$ parameter, \nova preferes the unitary mixing value $\alpha_{00}=1$ as the best fit value for both NH and IH. The combined analysis prefers $\alpha_{00}=0.97$ ($1$) as the best-fit point for NH (IH). T2K data alone prefer a best-fit of $\alpha_{00}$ closer to unitary, $\alpha_{00}=0.97$ for NH, and the unitary mixing case is allowed at $1\,\sigma$ C.L. All three cases allow large violation of unitary mixing at $90\%$ and $3\,\sigma$ C.L.  

In table~\ref{tab:bf_00}, we have enlisted the best-fit values of the unknown parameter for non-unitary mixing arising due to $\alpha_{00}$. It can be observed that all three cases, namely data from \nova, T2K, and the combined data from both the experiments, prefer unitary mixing or little deviation from unitary mixing as their best-fits. However, larger violation from unitary mixing is still allowed at $90\%$ limit. The $90\%$ limits obtained from the analysis of the \nova and T2K data are stronger than the limits obtained from their previous data in ref.~\cite{Miranda:2019ynh}. However these limits are still weaker compared to the global limits obtained in ref.~\cite{Blennow:2025qgd}.

For the $\alpha_{10}$ parameter, the results of both experiments are more consistent with each other. Both experiments allow each other's best-fit points for both hierarchies at $1\,\sigma$. For NH, T2K rules out the unitary mixing value $|\alpha_{10}|=0$ with a $\dchsq$ od $1.24$. For NH, both the experiments prefer best-fit point at $\alpha_{10}=0.06$. Hence, for NH, both experiment prefer a best-fit value closer to unitary mixing, as compared to the earlier best-fit value found in ref.~\cite{Miranda:2019ynh}. Therefore, the present $1\,\sigma$ tension for NH between the \nova and T2K data can be reduced with non-unitary mixing scheme, where non-unitary mixing arises due to $|\alpha_{10}|=0.06$. However, it is to be noted that this value of $|\alpha_{10}|$ is ruled out at $90\%$ C.L. by the global fit \cite{Blennow:2025qgd} of the only neutrino oscillation data.

In table~\ref{tab:bf_10}, we have enlisted the best-fit values for unknown parameters for non-unitary mixing arising due to $\alpha_{10}$. The $90\%$ limits of $|\alpha_{10}|$ are also given. It can be seen that for \nova data alone, the unitary mixing value of $|\alpha_{10}=0|$ falls withing the $1\,\sigma$ range. However, for T2K data alone, $|\alpha_{10}=0|$ falls outside $1\,\sigma$ range for IH. For the combined data, $|\alpha_{10}=0|$ falls outside $1\,\sigma$ range for both NH and IH. The $90\%$ limits on $|\alpha_{10}|$. in case of only \nova data and the combined data, are stronger than the previous limits obtained in ref.~\cite{Miranda:2019ynh}. However these limits are still weaker than the global limit obtained in ref.~\cite{Blennow:2025qgd}. 

In case of $\alpha_{11}$, the preference for unitary mixing is even stronger compared to that for $\alpha_{00}$. In table ~\ref{tab:bf_11}, we have enlisted the best-fit parameter values as well as the $90\%$ limits on $\alpha_{11}$. In this case also, the constraints on unitary violation are stronger than those in ref.~\cite{Miranda:2019ynh}, and weaker compared to the global fit constraints of ref.~\cite{Blennow:2025qgd}.

\begin{table*}[t]
\caption{Parameter values at the best-fit points for NO$\nu$A and T2K, when the non-unitary mixing arises due to $\alpha_{00}$. 
The $1\sigma$ errors are given where possible and $90\%$ C.L. limits for 1 degree of freedom (d.o.f.) are listed.}
\centering
\begin{ruledtabular}
\begin{tabular}{lcc|cc|}
 & \multicolumn{2}{c}{Best fit} & \multicolumn{2}{c}{$90\%$ C.L.} \\
 & NH & IH & NH & IH \\
\hline
\textit{Min.\ $\chi^2$ (degree of freedom)} \\
NO$\nu$A & 61.36 (53) &  &  &  \\
\\
T2K      & 94.48 (83) &  &  &  \\
\\
NO$\nu$A+T2K & 158.60 (141) &  &  &  \\
\hline
\textit{Min.\ $\Delta\chi^2$} \\
NO$\nu$A & 0 & 1.31 &  &  \\
\\
T2K      & 0 & 0.58 &  &  \\
\\
NO$\nu$A+T2K & 1.18 & 0 &  &  \\
\hline
$\sin^2\theta_{23}$ \\
NO$\nu$A & $0.57^{+0.01}_{-0.01}$ & $0.57^{+0.01}_{-0.02}$ & & \\
\\
T2K      & $0.56^{+0.02}_{-0.10}$ & $0.56^{+0.01}_{-0.01}$ & & \\
\\
NO$\nu$A+T2K & $0.57^{+0.01}_{-0.01}$ & $0.57^{+0.01}_{-0.01}$ & & \\
\hline
$\delta_{\rm CP}\ (^\circ)$ \\
NO$\nu$A & $(150^{+30}_{-140})$ & $-(90^{+40}_{-40})$ & & \\
\\
T2K      & $-(90^{+60}_{-50})$ & $-(90^{+30}_{-30})$ & & \\
\\
NO$\nu$A+T2K & $-(160^{+20}_{-60})$ & $-(100^{+20}_{-30})$ & & \\
\hline
$\alpha_{00}$ \\
NO$\nu$A & $1.00$ & $1.00$ & $>0.86$ & $>0.94$ \\
\\
T2K      & $0.97$ & $1.00$ & $>0.86$ & $>0.92$ \\
\\
NO$\nu$A+T2K & $0.97$ & $1.00$ & $>0.90$ & $>0.94$ \\
\end{tabular}
\end{ruledtabular}
\label{tab:bf_00}
\end{table*}

\begin{table*}[t]
\caption{Parameter values at the best-fit points for NO$\nu$A and T2K, when the non-unitary mixing arises due to $\alpha_{10}$. 
The $1\sigma$ errors are given where possible and $90\%$ C.L. limits for 1 degree of freedom (d.o.f.) are listed.}
\centering
\begin{ruledtabular}
\begin{tabular}{lcc|cc|}
 & \multicolumn{2}{c}{Best fit} & \multicolumn{2}{c}{$90\%$ C.L.} \\
 & NH & IH & NH & IH \\
\hline
\textit{Min.\ $\chi^2$ (degree of freedom)} \\
NO$\nu$A & 61.11 (53) &  &  &  \\
\\
T2K      & 93.57 (83) &  &  &  \\
\\
NO$\nu$A+T2K & 157.02 (141) &  &  &  \\
\hline
\textit{Min.\ $\Delta\chi^2$} \\
NO$\nu$A & 0 & 0.81 &  &  \\
\\
T2K      & 0.34 & 0 &  &  \\
\\
NO$\nu$A+T2K & 1.97 & 0 &  &  \\
\hline
$\sin^2\theta_{23}$ \\
NO$\nu$A & $0.46^{+0.01}_{-0.01}\oplus0.57^{+0.01}_{-0.03}$ & $0.46^{+0.01}_{-0.01}\oplus 0.57^{+0.01}_{-0.01}$ & & \\
\\
T2K      & $0.47^{+0.11}_{-0.03}$ & $0.46^{+0.04}_{-0.01}\oplus 0.55^{+0.02}_{-0.01}$ & & \\
\\
NO$\nu$A+T2K & $0.47^{+0.01}_{-0.01}\oplus 0.57^{+0.01}_{-0.01}$ & $0.46^{+0.01}_{-0.01}$ & & \\
\hline
$\delta_{\rm CP}\ (^\circ)$ \\
NO$\nu$A & $-(20^{+50}_{-200})$ & $-(80^{+50}_{-40})$ & & \\
\\
T2K      & $-(70^{+70}_{-70})$ & $-(120^{+50}_{-50})$ & & \\
\\
NO$\nu$A+T2K & $-(160^{+20}_{-30})$ & $-(90^{+40}_{-30})$ & & \\
\hline
$|\alpha_{10}|$ \\
NO$\nu$A & $0.06^{+0.05}_{-0.06}$ & $0.03^{+0.04}_{-0.03}$ & $<0.14$ & $<0.09$ \\
\\
T2K      & $0.06^{+0.09}_{-0.06}$ & $0.07^{+0.06}_{-0.05}$ & $<0.20$ & $<0.18$ \\
\\
NO$\nu$A+T2K & $0.04^{+0.04}_{-0.03}$ & $0.03^{+0.02}_{-0.01}$ & $<0.08$ & $<0.08$ \\
\end{tabular}
\end{ruledtabular}
\label{tab:bf_10}
\end{table*}

\begin{table*}[t]
\caption{Parameter values at the best-fit points for NO$\nu$A and T2K, when the non-unitary mixing arises due to $\alpha_{11}$. 
The $1\sigma$ errors are given where possible and $90\%$ C.L. limits for 1 degree of freedom (d.o.f.) are listed.}
\centering
\begin{ruledtabular}
\begin{tabular}{lcc|cc|}
 & \multicolumn{2}{c}{Best fit} & \multicolumn{2}{c}{$90\%$ C.L.} \\
 & NH & IH & NH & IH \\
\hline
\textit{Min.\ $\chi^2$ (degree of freedom)} \\
NO$\nu$A & 61.23 (53) &  &  &  \\
\\
T2K      & 94.70 (83) &  &  &  \\
\\
NO$\nu$A+T2K & 158.61 (141) &  &  &  \\
\hline
\textit{Min.\ $\Delta\chi^2$} \\
NO$\nu$A & 0 & 1.43 &  &  \\
\\
T2K      & 0 & 0.35 &  &  \\
\\
NO$\nu$A+T2K & 1.65 & 0 &  &  \\
\hline
$\sin^2\theta_{23}$ \\
NO$\nu$A & $0.57^{+0.01}_{-0.01}$ & $0.57^{+0.01}_{-0.01}$ & & \\
\\
T2K      & $0.57^{+0.04}_{-0.12}$ & $0.57^{+0.01}_{-0.01}$ & & \\
\\
NO$\nu$A+T2K & $0.57^{+0.01}_{-0.01}$ & $0.57^{+0.01}_{-0.01}$ & & \\
\hline
$\delta_{\rm CP}\ (^\circ)$ \\
NO$\nu$A & $(160^{+20}_{-110})$ & $-(90^{+40}_{-40})$ & & \\
\\
T2K      & $-(80^{+70}_{-50})$ & $-(90^{+30}_{-50})$ & & \\
\\
NO$\nu$A+T2K & $-(170^{+10}_{-20})$ & $-(100^{+20}_{-30})$ & & \\
\hline
$\alpha_{11}$ \\
NO$\nu$A & $0.99$ & $1.00$ & $>0.96$ & $>0.97$ \\
\\
T2K      & $0.99$ & $1.00$ & $>0.95$ & $>0.96$ \\
\\
NO$\nu$A+T2K & $1.00$ & $1.00$ & $>0.97$ & $>0.97$ \\
\end{tabular}
\end{ruledtabular}
\label{tab:bf_11}
\end{table*}

We now explain the origin of the tension between the present \nova and T2K data, and its possible resolution through non-unitary mixing induced by $\alpha_{10}$, in terms of the behaviour of the $\nu_\mu \to \nu_e$ and $\bar{\nu}_\mu \to \bar{\nu}_e$ appearance probabilities. Following the approach of ref.~\cite{Rahaman:2021zzm}, we define a benchmark point corresponding to vacuum oscillations with maximal $\tz$ and $\dcp=0$, which we denote as $000$. Deviations from this benchmark modify the appearance probability $\pme$, and we characterise these modifications in a qualitative way.

We use the symbols $+$ and $-$ to indicate whether a given choice of oscillation parameters enhances or suppresses $\pme$ relative to the benchmark point. Matter effects enhance (suppress) $\pme$ for normal (inverted) mass hierarchy, which we denote by $+$ ($-$). Similarly, placing $\theta_{23}$ in the higher (lower) octant enhances (suppresses) $\pme$, again denoted by $+$ ($-$). Finally, $\delta_{\rm CP}=-90^\circ$ ($+90^\circ$) enhances (suppresses) $\pme$ and is labelled by $+$ ($-$). It is important to note that while the effects of the mass hierarchy and $\delta_{\rm CP}$ reverse sign for $\pmebar$, the effect of the $\theta_{23}$ octant is the same for both neutrinos and antineutrinos. In table~\ref{parameter_label}, we have listed different parameters labels and their effects on $\pme$ and $\pmebar$.

\begin{table}[htbp!]
\hspace*{-1cm}
\begin{tabular}{|c|c|c|c|c|}
  \hline
  Parameter & Parameter value &
  Label & Effects on $\pme$ & Effects on $\pmebar$ \\
  \hline
  Hierarchy & Vacuum & $0$ & Benchmark & Benchmark \\
  \hline
  Hierarchy & NH & $+$ & Boost & Suppress \\
  \hline
  Hierarchy & IH & $-$ & Suppress & Boost \\
  \hline
  $\dcp$ & $0$ & $0$ & Benchmark & Benchmark \\
  \hline
  $\dcp$ & $-90^\circ$ & $+$ & Boost & Suppress \\
  \hline
  $\dcp$ & $+90^\circ$ & $-$ & Suppress & Boost \\
  \hline
  $\sin^2\tz$ & $0.5$ & $0$ & Benchmark & Benchmark \\
  \hline
  $\sin^2\tz$ & $>0.5$ & $+$ & Boost & Boost \\
  \hline
  $\sin^2\tz$ & $<0.5$ & $-$ & Suppress & Suppress \\
  \hline
\end{tabular}
\caption{Labels for different parameter values and their effects on oscillation probabilities.}
\label{parameter_label}
\end{table}

At the benchmark point $000$, the expected (signal + background) event numbers for \nova are $170$ for $\nu_e$ and $33$ for $\bar{\nu}_e$, while the observed numbers are $181$ and $32$, respectively. This indicates a moderate enhancement in the $\nu_e$ appearance channel relative to the benchmark expectation. Within the standard unitary mixing framework, such a moderate enhancement can arise from several competing effects. In particular, the combinations $++-$, $+-+$, and $-++$ can reproduce the observed $\nu_e$ excess. Among these, the $\nu_e$ data favour the $++-$ and $-++$ configurations.

For the $\bar{\nu}_e$ channel, the observed event rate is consistent with the benchmark expectation. Due to the limited antineutrino statistics, most parameter combinations remain allowed. The only exceptions are $+-+$ and $-+-$, which correspond to the minimum and maximum expected $\bar{\nu}_e$ event rates, respectively. Taken together, the unitary-mixing analysis of the \nova data therefore favours solutions of the form $++-$ and $-++$.

In case of T2K, at the benchmark point $000$, the expected (signal + background) event numbers are $79$ for $\nu_e$ and $19$ for $\bar{\nu}_e$, while the observed numbers are $107$ and $15$, respectively. This indicates a large enhancement in the $\nu_e$ appearance channel relative to the benchmark expectation. Within the standard unitary mixing framework, such a large enhancement can arise from the combination $+++$. The combination $-++$ is also allowed at $1\,\sigma$.

For the $\bar{\nu}_e$ channel, the observed event rate is moderately suppressed from the benchmark expectation. Due to the limited antineutrino statistics, all parameter combinations remain allowed. Taken together, the unitary-mixing analysis of the T2K data therefore favours solutions of the form $+++$ and $-++$.

When non-unitary mixing due to $\alpha_{10}$ is included, the situation changes qualitatively. In this case, $\alpha_{10}$ induces a correlated modification of both $\pme$ and $\pmebar$: for $\delta_{\rm CP}=-90^\circ$ ($+90^\circ$), both $\nu_e$ and $\bar{\nu}_e$ event rates are enhanced (suppressed). At the \nova best-fit region, for non-unitary mixing scheme due to $\alpha_{10}$, labelled by $++0$, the expected event numbers are $249$ for $\nu_e$ and $34$ for $\bar{\nu}_e$. However, a near-degenerate solution exists at $--+$, with $197$ $\nu_e$ and $38$ $\bar{\nu}_e$ events. The benchmark point $000$ also becomes viable once $\alpha_{10}$ is allowed. Although the $+++$ configuration predicts a $\nu_e$ rate significantly larger than observed, it reproduces the observed $\bar{\nu}_e$ event number almost exactly, and is therefore allowed at the $1,\sigma$ level. The $+--$ configuration is similarly allowed at $1,\sigma$.

For T2K, the best-fit point, for non-unitary mixing scheme due to $\alpha_{10}$, corresponds to the $--+$ configuration, yielding $108$ $\nu_e$ and $19$ $\bar{\nu}_e$ events, in excellent agreement with the observed values of $107$ and $15$. A near-degenerate solution is found for $+-+$, with $114$ $\nu_e$ and $17$ $\bar{\nu}_e$ events. At the standard best-fit point $+++$, the predicted $\nu_e$ rate ($133$ events) significantly exceeds the observed value; however, the predicted $\bar{\nu}_e$ rate ($21$ events) remains close to the observed one, allowing this configuration at the $1,\sigma$ level. The $-++$ configuration is also allowed at $1,\sigma$.

A detailed discussion of the impact of $\alpha_{10}$ on oscillation probabilities and appearance event rates is presented in Appendix~\ref{prob+event}. From this analysis, we conclude that the tension between \nova and T2K can be alleviated by non-unitary mixing driven by $\alpha_{10}$, with a preferred value $|\alpha_{10}|=0.06$. We emphasise, however, that this value lies outside the current $90\%$ confidence-level bound from global fits~\cite{Blennow:2025qgd}.

\begin{figure}[h!]
\centering
\includegraphics[width=0.4\textwidth] {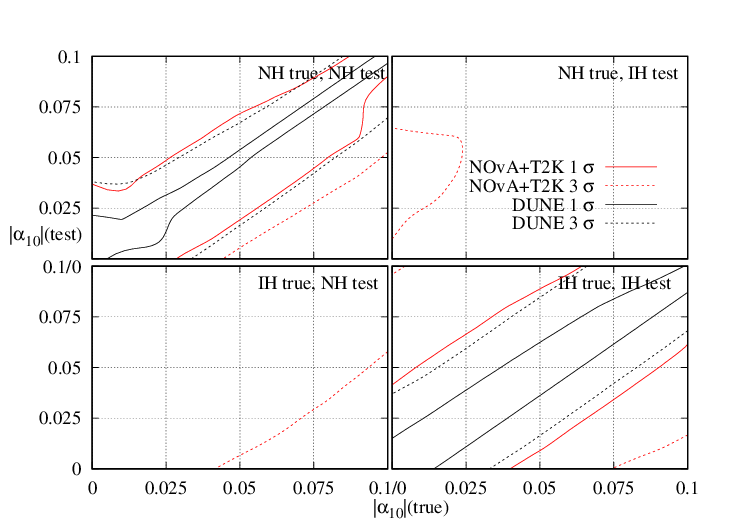}
\includegraphics[width=0.4\textwidth] {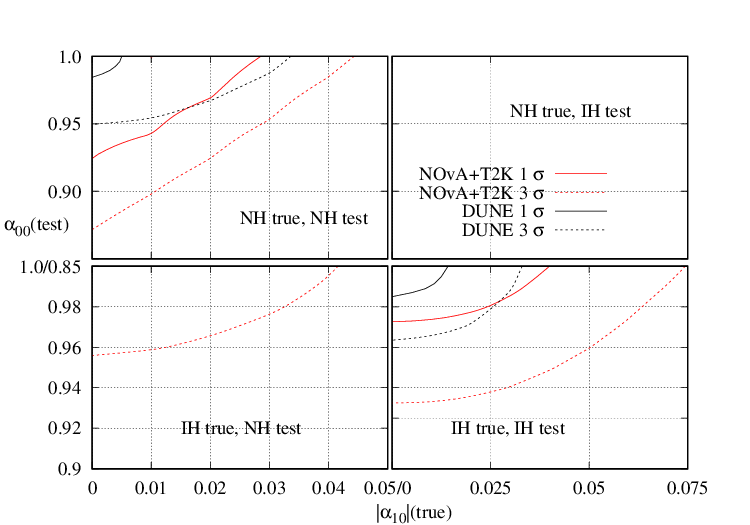}
\includegraphics[width=0.4\textwidth] {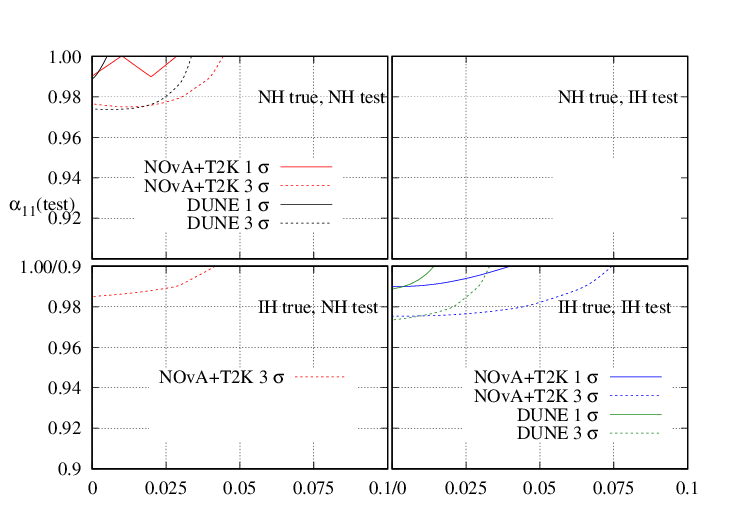}
\caption{\footnotesize{Sensitivity of $|\alpha_{10}|$, $\alpha_{00}$ and $\alpha_{11}$ assuming $\alpha_{10}$ as the true parameter for future combination of \nova and T2K, and DUNE.}}
\label{future}
\end{figure}

\section{Future sensitivity} 
\label{ch:future}
We have computed the sensitivity of $\alpha_{00}$ and $\alpha_{10}$ in the form of contour plots assuming $\alpha_{10}$ as the true parameter value. We have considered a combination of future \nova results with $13.305\times 10^{21}$ ($6.25\times 10^{21}$) POTs collected for a $\nu$ ($\bar{\nu}$) run along with future T2K results with $9.85\times10^{21}$ ($8.15\times 10^{21}$) POTs collected for a $\nu$ ($\bar{\nu}$) run. We have also separately considered DUNE with a $\nu$ and $\anu$ run, each corresponding to $5.5\times10^{21}$ POTs collected. We have presented the result in the form of contour plots in fig.~\ref{future} with true values $|\alpha_{10}|$ on the x-axis and the test values of $|\alpha_{10}|$ and $\alpha_{00}$ on the y-axis. To generate these plots, we fixed the true values of standard oscillation parameters at their current global best-fit values given in ref.~\cite{Esteban:2024eli}. The true values of $|\alpha_{10}|$ have been varied in the range $[0:0.1]$, with true $\phi_{10}=0$. For test parameters, we varied $\dcp$ in its complete range, while $\sin^2\tz$ and $|\dl|$ have been varied in their current $3\sigma$ range given in ref.~\cite{Esteban:2024eli}. Other standard parameters' test values have been fixed to their best-fit values. For non-unitary parameters, we varied the test values of $|\alpha_{10}|$ in the range $[0:0.1]$ and test values of $\phi_{10}$ in the range $[-180^\circ:180^\circ]$. We marginalised the $\dchsq$ over all the test parameters except $|\alpha_{10}|$. When $\alpha_{00}$ ($\alpha_{11}$) is the test parameter, we varied it in the range $[0.7:1]$ and marginalised $\dchsq$ over the standard test parameters. 

It can be seen from fig.~\ref{future} that when non-unitary mixing arises due to $\alpha_{10}$, and when true and test hierarchies are the same, the test values of $|\alpha_{10}|$ can be ruled out at $1\,\sigma$ outside the range of the true values within a $\pm 0.03$ uncertainty  by the combination of future \nova and T2K data. A future DUNE run can exclude the test values of $|\alpha_{10}|$ outside the range of true value within a $\pm 0.01$ uncertainty. When true and test hierarchies are opposite, then the combination of \nova and T2K rules out regions outside $0\leq\alpha_{10}({\rm true})\leq0.025$ ($0.045<\alpha_{10}({\rm true})\leq0.1$) and $0\leq\alpha_{10}({\rm test})\leq0.063$ ($0<\alpha_{10}({\rm test})\leq0.06$) for NH true-IH test (IH true-NH test) at $3\,\sigma$ C.L. DUNE rules out the wrong hierarchy at a $3\,\sigma$ level. 

When true and test hierarchies are the same, the combination of a \nova and T2K future run allows for a very small region corresponding to $0\leq|\alpha_{10}| ({\rm true})\leq0.025$ ($0\leq|\alpha_{10}|({\rm true})\leq0.045$) and $0.92\leq \alpha_{00}({\rm test})\leq 1$ ($0.87\leq \alpha_{00}({\rm test})\leq 1$) at $1\,\sigma$ ($3\,\sigma$) C.L. The future DUNE run allows for a tiny region close to $|\alpha_{10}| ({\rm true})=0$ and $\alpha_{00}({\rm test})=1$ at a $1\,\sigma$ C.L. At $3\,\sigma$, DUNE allows for $0\leq|\alpha_{10}| ({\rm true})\leq0.03$ and test $0.95\leq \alpha_{00}({\rm test})\leq 1$. When NH is the true hierarchy, the future combination of \nova and T2K results, as well as DUNE can rule out an IH test at $3\,\sigma$ level, for a $\alpha_{00} ({\rm test})$. When IH is the true hierarchy, the combination of \nova and T2K results rule out the NH test outside the range $0\leq|\alpha_{10}| ({\rm true})\leq0.04$ and $0.95\leq \alpha_{00}({\rm test})\leq 1$ at $3\,\sigma$. DUNE rules out the NH test completely at $3\,\sigma$.

In case of $\alpha_{11}$ test, the combination of future \nova and T2K, as well as DUNE, have stronger exclusion potential to rule out $\alpha_{11}$.

\section{Conclusion} 
\label{ch:conclusion}
In case of non-unitary mixing due to $\alpha_{00}$, and $\alpha_{11}$, data from both \nova and T2K prefer unitary mixing or very little deviation from unitary mixing as their best-fit solutions. The $90\%$ limits provided by these experiments are stronger than the previous limits obtained from these experiments. However, these constraints are still weaker comparable to the constrained provided by the present global-fit. When unitary violation arises due to $\alpha_{10}$, both the experiments, as well as their combined analysis prefer slightly larger unitary violation as their best-fit solution. For \nova and the combined analysis, the $90\%$ limits on $|\alpha_{10}|$ are stronger than before.

The tension between \nova and T2K arises from the $\nu_e$ appearance channel. \nova observed a moderate excess in its electron appearance event numbered compared to the expected event numbers for the benchmark parameter values, namely vacuum oscillation, $\tz$ maximal and $\dcp=0$. This moderate excess can be accommodated with the combination of NH, $\tz$ in HO, $0<\dcp<180^\circ$ and IH, $\tz$ in HO, and $-180^\circ<\dcp<0$. On the other hand, T2K observes a large excess in the observed electron event numbers, compared to the benchmark point. This large excess can only be accommodated with $\dcp$ firmly anchored around $-90^\circ$. This gives rise to the tension at NH. A combination of the two experiments prefers IH over NH. In the case of $\alpha_{10}$ being the reason for non-unitary mixing, the $\nu_e$ appearance events of both the experiments see a boost (suppression) for $\dcp=-90^\circ$ ($90^\circ$) for both the hierarchies and octants of $\tz$. Thus, in this case, $\tz$ in LO becomes a viable solution for both experiments. In this case, both experiments have large overlap between the allowed regions at $1\,\sigma$ on the $\sin^2\tz-\dcp$ plane. Both experiment have a preference for non-unitary mixing with best-fit point at $|\alpha_{10}|= 0.06$ for NH. The future run of \nova and T2K have good potential to rule out the wrong values of $|\alpha_{10}|$ as well as $\alpha_{00}$ and $\alpha_{11}$ if non-unitary mixing arises due to $\alpha_{10}$. The sensitivity is improved by future DUNE data.

\bibliographystyle{apsrev}
\bibliography{referenceslist}
\appendix
\section{Analysis details}
\label{analysis}
The T2K experiment \cite{Itow:2001ee} uses the $\nu_\mu$ beam from the J-PARC accelerator at Tokai and the water Cerenkov detector at Super-Kamiokande, which is
295 km away from the source. The detector is situated
$2.5^\circ$ off-axis. The flux peaks at $0.7$ GeV, which is also
close to the first oscillation maximum. T2K started taking data in 2009 and up until 2020 released results \cite{T2Kapp, T2Kdisapp} corresponding to $1.97\times 10^{21}$ ($1.63\times 10^{21}$) protons on target (POTs) in neutrino (anti-neutrino) mode. 

The \nova detector \cite{nova_tdr} is a 14 kt totally active scintillator detector (TASD), placed 810 km away from the neutrino source at Fermilab, situated $0.8^\circ$ off-axis with respect to the NuMI beam. The flux peaks at 2 GeV,
close to the oscillation maxima at 1.4 GeV (1.8 GeV) for NH (IH). \nova started taking data in 2014 and as of the 2024 data release \cite{Wolcott:2024}, has collected $2.661\times10^{21}$ ($1.250\times10^{21}$) POTss, for neutrino (anti-neutrino) mode. 

Since the T2K data are from 2020, in order to analyze the data from both of the experiments, we have used
the 2019 global-best fit values for standard oscillation parameters \cite{Esteban:2018azc}. We have fixed $\ds$ and $\tx$ to their best-fit values. The values of $\sin^2\ty$, $\sin^2\tz$ and $\Delta_{3l}$, with $l=1$ ($2$) for NH (IH) have been varied in their $3\,\sigma$ range. $\dcp$ has been varied in its complete range $[-180^\circ:180^\circ]$. Among the non-unitary parameters, $\alpha_{00}$ and $\alpha_{11}$ have been varied within the range $[0.7:1.0]$, while $|\alpha_{10}|$ has been varied within the range $[0:0.3]$, and $\phi_{10}$ has been allowed to take on any value $[-180^\circ:180^\circ]$. We have chosen these ranges to cover the $3\,\sigma$ regions given in ref.~\cite{Miranda:2019ynh}. We have used GLoBES \cite{Huber:2004ka} to calculate the theoretical event rates as well as the $\chi^2$ between theoretical event rates and experimental data. To do so, we fixed the bin based detector efficiencies by matching with the simulated event numbers provided by \nova \cite{Wolcott:2024} and T2K collaborations \cite{T2Kapp, T2Kdisapp}. For energy resolution, we used a Gaussian function
\begin{equation}
R^c (E,E^\prime)=\frac{1}{\sqrt{2\pi}}e^{-\frac{(E-E^\prime)^2}{2\sigma^2(E)}},
\end{equation}
where $E^\prime$ is the reconstructed energy. The energy resolution function is given by 
\begin{equation}
\sigma(E)=\alpha E+\beta \sqrt{E}+\gamma,
\label{res}
\end{equation}
where $\alpha=0$, $\beta=0.075$, $\gamma=0.05$ for T2K. For NO$\nu$A, however, we used $\alpha=0.11$ ($0.09$), $\beta=\gamma=0$ for $\nu_e$ ($\nu_\mu$) events. For systematics uncertainty, we have used $5\%$ energy calibration and flux normalization backgrounds for both of the experiments. The experimental event rates have been taken from ref.~\cite{T2Kapp, T2Kdisapp} for T2K, and \cite{Wolcott:2024} for NO$\nu$A.
\section{Analysis of \nova and T2K data with unitary mixing scheme}
\label{uni-analysis}
\begin{figure}[h]
\centering
\includegraphics[width=0.5\textwidth] {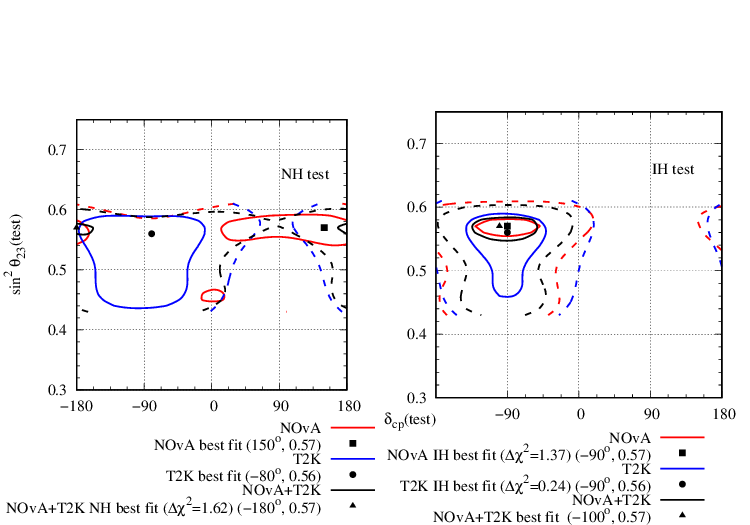}
\caption{\footnotesize{Allowed regions in the $\sin^2\tz-\dcp$ plane for \nova and T2K after analysing the data with standard unitary mixing. The left (right) panel is for NH (IH). The red (blue) line indicates \nova (T2K), and the black line indicates the combined data. The solid (dotted) lines indicate the boundaries of the $1\,\sigma$ ($3\,\sigma$) allowed regions. }}
\label{result-uni}
\end{figure}

In this section, we present the analysis, with standard unitary mixing scheme, of \nova and T2K latest data. From fig.~\ref{result-uni}, it can be seen that the best-fit points of the two experiments are far apart from each other. There are no overlaps between the $1\,\sigma$ allowed regions of the two experiments for NH. Both experiments have their best-fit points at NH. However, T2K has a near degenerate best-fit point at IH. The combined analysis prefers IH over NH. Only a small area near the $\dcp$ conserving values at NH are allowed at $1\,\sigma$. These results are in agreements with the results reported by the joint analysis of \nova and T2K collaborations \cite{T2K:2025wet}.
\section{Oscillation probabilities and event numbers of \nova and T2K}
\label{prob+event}
In this section, we will discuss the effect of non-unitary mixing due $\alpha_{10}$ on oscillation probabilities $\pme$ and $\pmebar$ as well as the $\nu_e$ and $\bar{\nu}_e$ event numbers.

\begin{figure}[h!]
\centering
\includegraphics[width=0.5\textwidth] {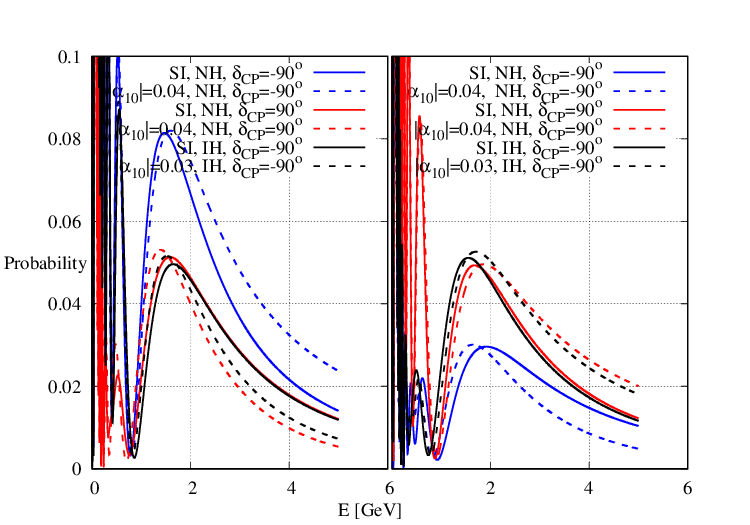}
\includegraphics[width=0.5\textwidth] {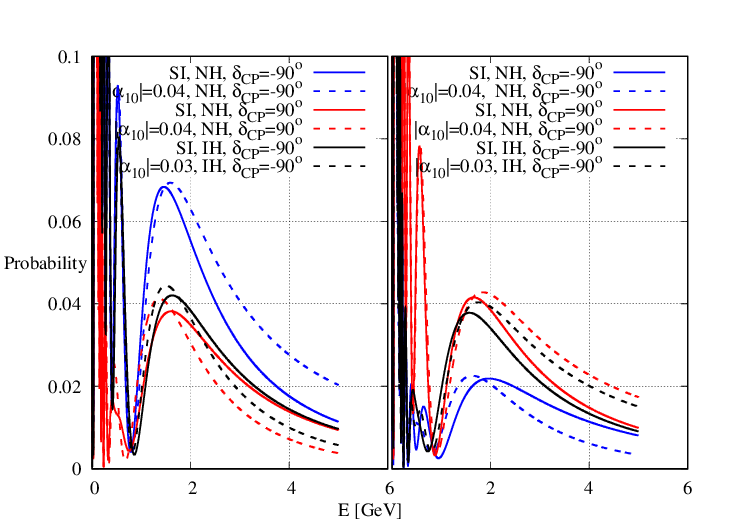}
\caption{\footnotesize{$\nu_\mu \to \nu_e$ (left panel) and $\bar{\nu}_\mu \to \bar{\nu}_e$ (right panel) oscillation probability as a function of energy with different hierarchy-$\dcp$ combinations for standard oscillation and non-unitary mixing due to $\alpha_{10}$ for the \nova experiment. The oscillation parameter values including $|\alpha_{10}|$ are fixed to the combined best-fit values of \nova and T2K. For NH (IH), $\phi_{10}=120^\circ$ ($60^\circ$). The left (right) panels are for neutrino (anti-neutrino) probabilities, and the top (bottom) panels are for $\tz$ in HO (LO). For HO (LO), we have used $\sin^2\tz=0.57$ (0.43).}}
\label{prob-alpha10-nova}
\end{figure}

\begin{figure}[h!]
\centering
\includegraphics[width=0.5\textwidth] {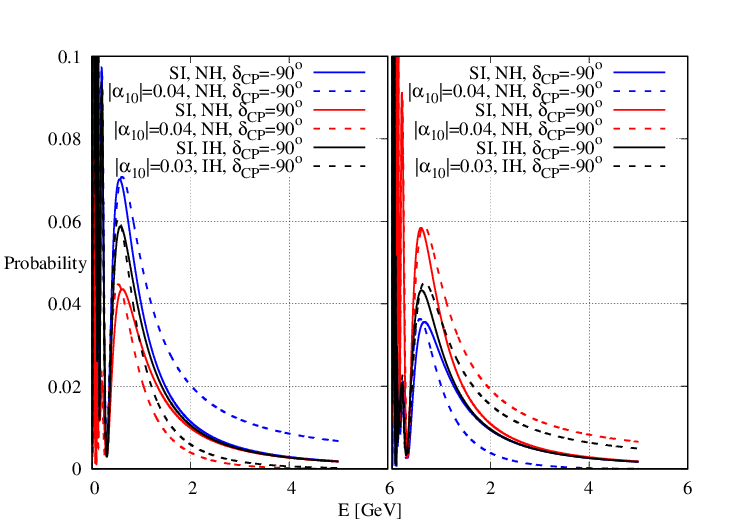}
\includegraphics[width=0.5\textwidth] {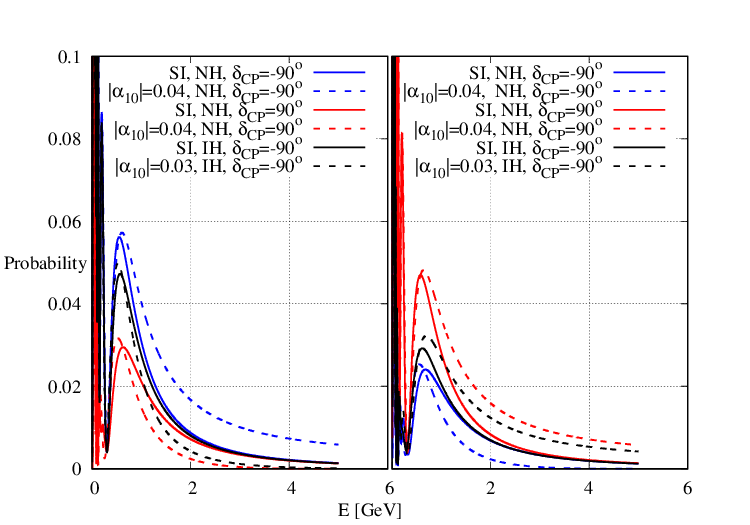}
\caption{\footnotesize{$\nu_\mu \to \nu_e$ (left panel) and $\bar{\nu}_\mu \to \bar{\nu}_e$ (right panel) oscillation probability as a function of energy with different hierarchy-$\dcp$ combinations for standard oscillation and non-unitary mixing due to $\alpha_{10}$ for the T2K experiment. The oscillation parameter values including $|\alpha_{10}|$ are fixed to the combined best-fit values of \nova and T2K. For NH (IH), $\phi_{10}=120^\circ$ ($60^\circ$). The left (right) panels are for neutrino (anti-neutrino) probabilities, and the top (bottom) panels are for $\tz$ in HO (LO). For HO (LO), we have used $\sin^2\tz=0.57$ (0.43).}}
\label{prob-alpha10-t2k}
\end{figure}

In fig.~\ref{prob-alpha10-nova}, we have shown $\pme$ and $\pmebar$ as a function of energy for \nova experiment and for different hierarchy-$\dcp$ combinations. The left (right) panels are for neutrino (anti-neutrino), and the top (bottom) panels are for $\tz$ in HO (LO). We have used $\sin^2\tz=0.57$ and $0.43$ for HO and LO respectively. Other parameters including $|\alpha_{10}|$ and $\phi_{10}$ have been fixed at the combined best-fit points of \nova and T2K. As can be seen, in case of non-unitary mixing due to $\alpha_{10}$, both $\pme$ and $\pmebar$ gets a slight boost at the oscillation peak energy compared to probabilities due to standard unitary mixing. However, for NH-$\dcp=90^\circ$ and IH-$\dcp=-90^\circ$, $\pme$ gets a moderate suppression after the oscillation maximum energy compared to the oscillation probabilities due to unitary mixing. In case of anti-neutrino, this suppression after the oscillation maximum energy takes place in case of NH-$\dcp=-90^\circ$. This feature remains same for both the octants of $\tz$. In fig.~\ref{prob-alpha10-t2k}, we have shown the similar probability plots for T2K experiment, and we can see the similar features for T2K as well.

We next examine how the expected total (signal + background) appearance event numbers for $\nu_e$ and $\bar{\nu}_e$ change as the oscillation parameters deviate from the benchmark values corresponding to vacuum oscillations with $\sin^2\theta_{23}=0.5$ and $\delta_{\rm CP}=0$, which we label as $000$. Table~\ref{nova_events} summarises the expected event numbers for the current \nova exposure.

At the benchmark point $000$, the expected event numbers for \nova are $170$ for $\nu_e$ appearance and $33$ for $\bar{\nu}_e$ appearance in the standard unitary mixing scenario, while the observed event numbers are $181$ and $33$, respectively. Thus, within unitary mixing, the benchmark point provides a good description of the $\bar{\nu}_e$ data but fails to account for the moderate excess observed in the $\nu_e$ channel.

When non-unitary mixing induced by $\alpha_{10}$ is included, the oscillation probabilities are enhanced, as discussed earlier. As a result, the expected event numbers at $000$ increase to $198$ for $\nu_e$ and $35$ for $\bar{\nu}_e$. Consequently, in the presence of $\alpha_{10}$, the benchmark point $000$ provides an acceptable fit within $1,\sigma$ for both $\nu_e$ and $\bar{\nu}_e$ appearance channels. In table~\ref{nova_events}, the expected event numbers corresponding to non-unitary mixing due to $\alpha_{10}$ are shown in square brackets.

We then vary one oscillation parameter at a time and compute the corresponding expected $\nu_e$ and $\bar{\nu}_e$ event numbers. Following the notation introduced in the main text, we find that, in the unitary mixing case, the parameter combinations $++-$ and $-++$ provide the closest agreement with the observed $\nu_e$ appearance data. Due to limited antineutrino statistics, all parameter combinations except $+-+$ and $-+-$ yield acceptable fits to the $\bar{\nu}_e$ data at the $1,\sigma$ level. These two excluded combinations correspond to the minimum and maximum expected $\bar{\nu}_e$ event rates for NO$\nu$A in the unitary scenario.

In the presence of non-unitary mixing due to $\alpha_{10}$, both $\nu_e$ and $\bar{\nu}_e$ appearance event numbers are enhanced (suppressed) for $\delta{\rm CP}=-90^\circ$ ($+90^\circ$), independently of the mass hierarchy and $\theta_{23}$ octant. At the benchmark point $000$, this enhancement leads to expected event numbers of $198$ for $\nu_e$ and $35$ for $\bar{\nu}_e$, making $000$ a viable solution at the $1,\sigma$ level. Additional solutions allowed at $1,\sigma$ for the $\nu_e$ channel include $++-$ and $--+$. For the $\bar{\nu}_e$ channel, all parameter combinations except $-++$ are allowed at $1,\sigma$. Combining the appearance and disappearance data, the $1,\sigma$ allowed regions correspond to the parameter labels $+++$, $++-$, $+--$, $+-0$, and $--+$. A small region around $-++$ is also allowed at $1,\sigma$ for two degrees of freedom.

For T2K, table~\ref{t2k_events} shows that, at the benchmark point $000$, the expected event numbers are $79$ for $\nu_e$ appearance and $19$ for $\bar{\nu}_e$ appearance, while the observed numbers are $107$ and $15$, respectively. This indicates a large enhancement in the $\nu_e$ channel and a moderate suppression in the $\bar{\nu}_e$ channel relative to the benchmark expectation. Within the unitary mixing framework, such a large enhancement in $\nu_e$ appearance can only be achieved when $\delta_{\rm CP}$ is close to $-90^\circ$, corresponding to the $+++$ configuration. The $-++$ configuration is also allowed at the $1,\sigma$ level.

When non-unitary mixing due to $\alpha_{10}$ is considered, the expected $\nu_e$ appearance event number at $000$ increases to $92$, making the benchmark point viable at $1,\sigma$. The best agreement with the observed $\nu_e$ data is obtained for the $++0$ and $--+$ configurations, which predict $107$ and $108$ $\nu_e$ events, respectively. Notably, the $--+$ configuration also provides an acceptable description of the NO$\nu$A $\nu_e$ appearance data. Other configurations allowed at $1,\sigma$ include $+00$ and $+-+$. For the $\bar{\nu}_e$ channel, all parameter combinations remain allowed at $1,\sigma$. The combined analysis therefore favours $--+$ as the new best-fit solution, with the $1,\sigma$ allowed regions given by $-++$, $+-+$, $+++$, and $++0$.

\begin{figure}[h]
\centering
\includegraphics[width=0.5\textwidth]{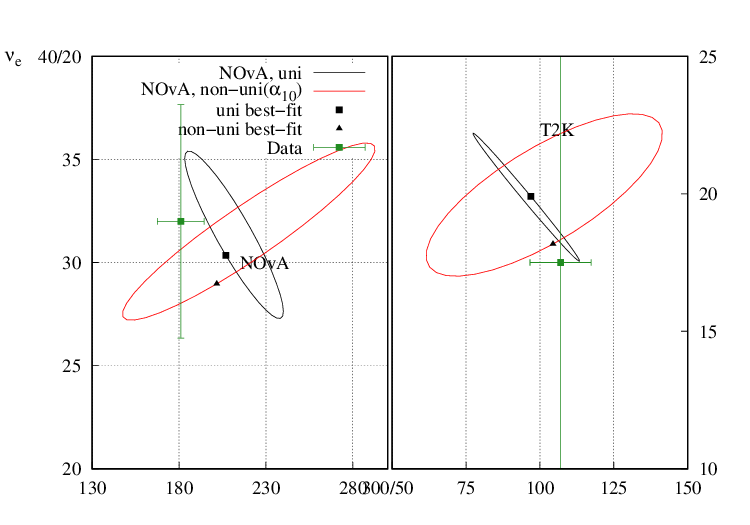}
\caption{\footnotesize{Bi-event plots for \nova (left) and T2K (right). $\dcp$ has been varied in the range $[-180^\circ:180^\circ]$. All other parameters have been fixed at the best-fit values for NH of the combined analysis. The black ellipse marks the case for Standard unitary mixing, while the red ellipse signifies the non-unitary mixing due to $\alpha_{10}$. The indicated best-fit points on the plot denote the best-fit point of the combined analysis. }}
\label{bievent}
\end{figure}

Finally, we illustrate these results using bi-event plots shown in fig.~\ref{bievent}. The expected $\nu_e$ and $\bar{\nu}_e$ appearance event numbers (signal + background) are computed for the current \nova and T2K exposures by varying $\delta_{\rm CP}$ in the range $[-180^\circ:180^\circ]$, while fixing all other oscillation parameters at the NH best-fit values of the combined analysis. The resulting $\bar{\nu}_e$ versus $\nu_e$ distributions form elliptical contours. In fig.~\ref{bievent}, the left (right) panel corresponds to \nova (T2K). The black ellipses represent standard unitary mixing, while the red ellipses correspond to non-unitary mixing induced by $\alpha_{10}$. The marked points indicate the combined best-fit values.

When non-unitary mixing arises due to $\alpha_{10}$, parts of the bi-event contours for both experiments move closer to the observed event numbers. Moreover, at the combined best-fit point, the predicted $\nu_e$ event number for \nova and both $\nu_e$ and $\bar{\nu}_e$ event numbers for T2K are closer to the data than in the unitary mixing case. This further supports our conclusion that the tension between \nova and T2K can be alleviated by non-unitary mixing driven by $\alpha_{10}$.
\begin{table}[htbp]
  \begin{center}
\begin{tabular}{|c|c|c|c|}
  \hline
  Hierarchy-$\sin^2\tz$-$\dcp$ & Label&
  $\nu_e$ Appearance&$\bar{\nu}_e$ Appearance \\
  &&events&events\\
  \hline
   Vacuum-$0.5$-$0$ & $000$ & $170.18$ & $32.97$ \\
   &&$[197.65]$&$[34.77]$\\
  \hline
  NH-$0.5$-$0$ & $+00$ & $194.11$& $28.72$\\
  &&$[225.90]$&$[30.87]$\\
  \hline
  NH-$0.57$-$0$  & $++0$ & $216.40$ & $32.01$ \\
  &&$[249.16]$&$[34.26]$\\
  \hline
  NH-$0.43$-$0$  & $+-0$ & $186.65$ & $27.78$ \\
  &&$[217.25]$&$[29.82]$\\
  \hline
 NH-$0.57$-$-90^\circ$ & $+++$ & $240.50$ & $27.05$\\
 &&$[268.88]$&$[32.17]$\\
 \hline
  NH-$0.57$-$+90^\circ$ & $++-$ & $183.98$ & $34.98$\\
  &&$[165.16]$&$[30.52]$\\
 \hline
 NH-$0.43$-$-90^\circ$  & $+-+$ & $210.43$ & $22.84$\\
 &&$[239.25]$&$[27.70]$\\
 \hline
 NH-$0.43$-$+90^\circ$  & $+--$ & $153.91$ & $30.77$ \\
 &&$[136.89]$&$[26.47]$\\
 \hline
 
 IH-$0.57$-$-90^\circ$  & $-++$ & $182.61$ & $34.94$ \\
 &&$[216.10]$&$[44.50]$\\
 \hline
  IH-$0.43$-$-90^\circ$  & $--+$ & $163.56$ & $28.97$ \\
  &&$[197.14]$&$[37.52]$\\
 \hline
 IH-$0.57$-$+90^\circ$  & $-+-$ & $138.47$ & $44.92$ \\
 &&$[121.64]$&$[35.37]$\\
 \hline
 IH-$0.43$-$+90^\circ$  & $---$ & $119.42$ & $38.96$ \\
 &&$[104.42]$&$[30.35]$\\
 \hline
\end{tabular}
\end{center}
 \caption{Expected $\nu_e$ and $\bar{\nu}_e$ appearance events of \nova for $2.661\times10^{21}$ ($1.25\times 10^{21}$) POTs in $\nu$ ($\bar{\nu}$) mode and for different combinations of the unknown parameter values for unitary mixing and non-unitary mixing. The expected event numbers for non-unitary mixing due to $|\alpha_{10}|=0.03$ ($0.04$) and $\phi_{10}=120^\circ$ ($60^\circ$) for NH (IH) have been given inside $[]$. The observed numbers of $\nu_e$ and $\bar{\nu}_e$ events are 181 and 32 respectively.}
  \label{nova_events}
\end{table}

\begin{table}[htbp!]
  \begin{center}
\begin{tabular}{|c|c|c|c|}
  \hline
  Hierarchy-$\sin^2\tz$-$\dcp$ & Label&
  $\nu_e$ Appearance&$\bar{\nu}_e$ Appearance \\
  &&events&events\\
  \hline
   Vacuum-$0.5$-$0$ & $000$ & $79.43$ & $19.04$ \\
   &&$[91.67]$&$[20.71]$\\
  \hline
  NH-$0.5$-$0$ & $+00$ & $84.86$& $18.27 $\\
  &&$[97.87]$&$[19.97]$\\
  \hline
  NH-$0.57$-$0$  & $++0$ & $93.77$ & $19.85$ \\
  &&$[107.21]$&$[21.70]$\\
  \hline
  NH-$0.43$-$0$  & $+-0$ & $76.91$ & $16.69$ \\
  &&$[89.44]$&$[18.20]$\\
  \hline
 NH-$0.57$-$-90^\circ$ & $+++$ & $113.32$ & $17.55$ \\
 &&$[132.79]$&$[20.46]$\\
 \hline
  NH-$0.57$-$+90^\circ$ & $++-$ & $77.42$ & $22.20$ \\
  &&$[66.08]$&$[19.22]$\\
 \hline
 NH-$0.43$-$-90^\circ$  & $+-+$ & $96.45$ & $14.39$ \\
 &&$[114.47]$&$[16.88]$\\
 \hline
 NH-$0.43$-$+90^\circ$  & $+--$ & $60.55$ & $19.04$ \\
 &&$[51.55]$&$[16.36]$\\
 \hline
 
 IH-$0.57$-$-90^\circ$  & $-++$ & $98.87$ & $19.15$ \\
 &&$[123.25]$&$[23.73]$\\
 \hline
  IH-$0.43$-$-90^\circ$  & $--+$ & $85.50$ & $15.10$\\
  &&$[107.82]$&$[19.36]$\\
 \hline
 IH-$0.57$-$+90^\circ$  & $-+-$ & $66.25$ & $24.56$ \\
 &&$[54.32]$&$[19.99]$\\
 
 \hline
 IH-$0.43$-$+90^\circ$  & $---$ & $52.48$ & $20.91$ \\
 &&$[43.30]$&$[16.91]$\\
 
 \hline
\end{tabular}
\end{center}
 \caption{Expected $\nu_e$ and $\bar{\nu}_e$ appearance events of T2K for $1.97\times10^{21}$ ($1.63\times 10^{21}$) POTs in $\nu$ ($\bar{\nu}$) mode and for different combinations of the unknown parameter values for unitary mixing and non-unitary mixing. The expected event numbers for non-unitary mixing due to $|\alpha_{10}|=0.03$ ($0.04$) and $\phi_{10}=120^\circ$ ($60^\circ$) for NH (IH) have been given inside $[]$. The observed numbers of $\nu_e$ and $\bar{\nu}_e$ events are 107 and 15 respectively.}
  \label{t2k_events}
\end{table}

\end{document}